\documentclass[11pt,twoside]{article}
\usepackage{asp2014}

\aspSuppressVolSlug
\resetcounters

\begin{document}

\title{RASCAL: Towards automated spectral wavelength calibration}

\author{Josh~Veitch-Michaelis and Marco~C~Lam}
\affil{Astrophysics Research Institute, Liverpool John Moores University, IC2, LSP, 146 Brownlow Hill, Liverpool, Merseyside L3 5RF, UK\\ \email{j.l.veitchmichaelis@ljmu.ac.uk}}

\paperauthor{Josh~Veitch-Michaelis}{}{ORCID}{Astrophysics Research Institute}{}{Liverpool}{Merseyside}{Postal Code}{Country}
\paperauthor{Marco Lam}{}{ORCID_Or_Blank}{Astrophysics Research Institute}{Author2 Department}{City}{State/Province}{Postal Code}{Country}



  
\begin{abstract}
Wavelength calibration is a routine and critical part of any spectral workflow, but many astronomers still resort to matching detected peaks and emission lines by hand. We present RASCAL (RANSAC Assisted Spectral CALibration), a python library for automated wavelength calibration of astronomical spectrographs. RASCAL implements recent state-of-the-art methods for wavelength calibration and requires minimal input from a user. In this paper we discuss the implementation of the library and apply it to real-world calibration spectra.

\end{abstract}

\section{Introduction}

Wavelength calibration is a routine and critical part of any spectral workflow. This usually involves (semi-manually) matching peaks in an arc lamp spectrum to a catalogue of known emission lines~\citep{10.1111/j.1365-2966.2010.17403.x} and may take minutes, even for an experienced user if the spectrum is particularly crowded. Automated pipelines typically involve some kind of cross-correlation or template matching with an existing lamp spectrum and require stable instruments.

While many observatories have published reduction pipelines that involve wavelength calibration~\citep{2002AJ....123..485S,2012ascl.soft03003C, 2013ApJS..208....5N}, most are not readily transferable to other instruments. Libraries like PypeIt~\citep{j_xavier_prochaska_2019_3506873} also exist for template matching. There is growing need for an automated solution that is easily transferable and robust to system re-configuration e.g. grating position, lamp type. This would be particularly useful in sharing a single data reduction pipeline among a network of small telescope facilities when staffing for software development and maintenance is limited.

In order to address this, we have developed RASCAL (RANSAC-Assisted Spectral CALibration). RASCAL only requires an atlas of calibration lines, a list of peaks, and some information about the system. RASCAL has been developed for the ASPIRED program \cite{P10-46_adassxxix} and broadly follows the algorithm presented in \cite{Song:18}. We are not aware of a public implementation of this algorithm prior to this paper.

We are releasing RASCAL as open-source as a Python library that can be easily integrated in to astronomical pipelines. The original paper only presents results from commercial spectrometers, so we contribute an initial evaluation on real-world spectra from astronomical instruments. We also present some tweaks and improvements to the original algorithm that result in improved correspondence matching.

\section{Calibration challenges}

Given a set of detected peak locations in a spectrum ($P$ [px]) and an atlas of emission lines ($A$ [$\lambda$]), our task is to find a matching member of $A$ for every member of $P$. Once the correspondences between $P$ and $A$ have been established, these are used to fit a model $f(x, p) = x_{\lambda}$ where $p$ are model parameters, $x$ is a detector location in pixels and $x_{\lambda}$ is the corresponding wavelength.

The emission lines in the atlas are assumed to be exact and taken from standard tables e.g. from the National Institute of Standards and Technology~\citep[NIST][]{nist_asd:2019} which collates values from the literature. No assumption is made about the peak finding routine, but this could be via manual line detection or a library routine like \texttt{scipy.signal.find\_peaks}.

It is possible that some detected peaks are spurious or correspond to a line not in the atlas. Vice versa, it is possible that some atlas lines were not detected because they are outside the spectral range of the detector, too low in amplitude, and so on. In fact, in the general case, any member of P could correspond to any member of A.

There is likely to be noise in the peak finding routines. This can be, for example due to detector noise or quantisation (e.g. not using sub-pixel peak fitting). There may also be complications such as blended lines - detected peaks which correspond to multiple emission lines, such as unresolved doublets. Once peaks and wavelengths have been matched, the model fitting process is largely straightforward. It is important that robust fitting methods are used, otherwise a single incorrect match can significantly skew the final model parameters.

\begin{figure}
\centering
\begin{minipage}{.49\textwidth}
  \centering
  \includegraphics[width=\linewidth]{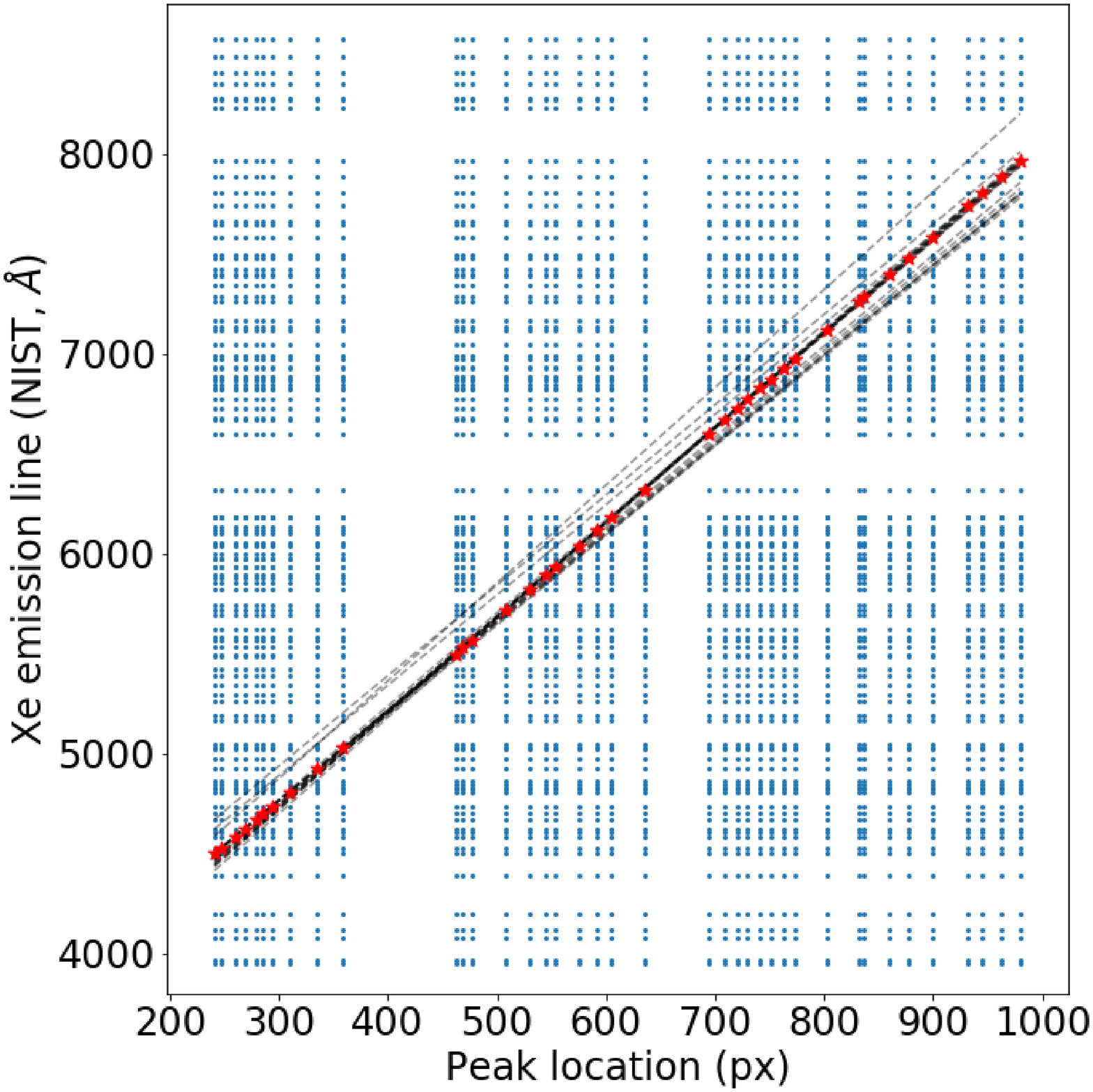}
    \label{fig:peaks}
\end{minipage}%
\begin{minipage}{.49\textwidth}
  \centering
    \includegraphics[width=\linewidth]{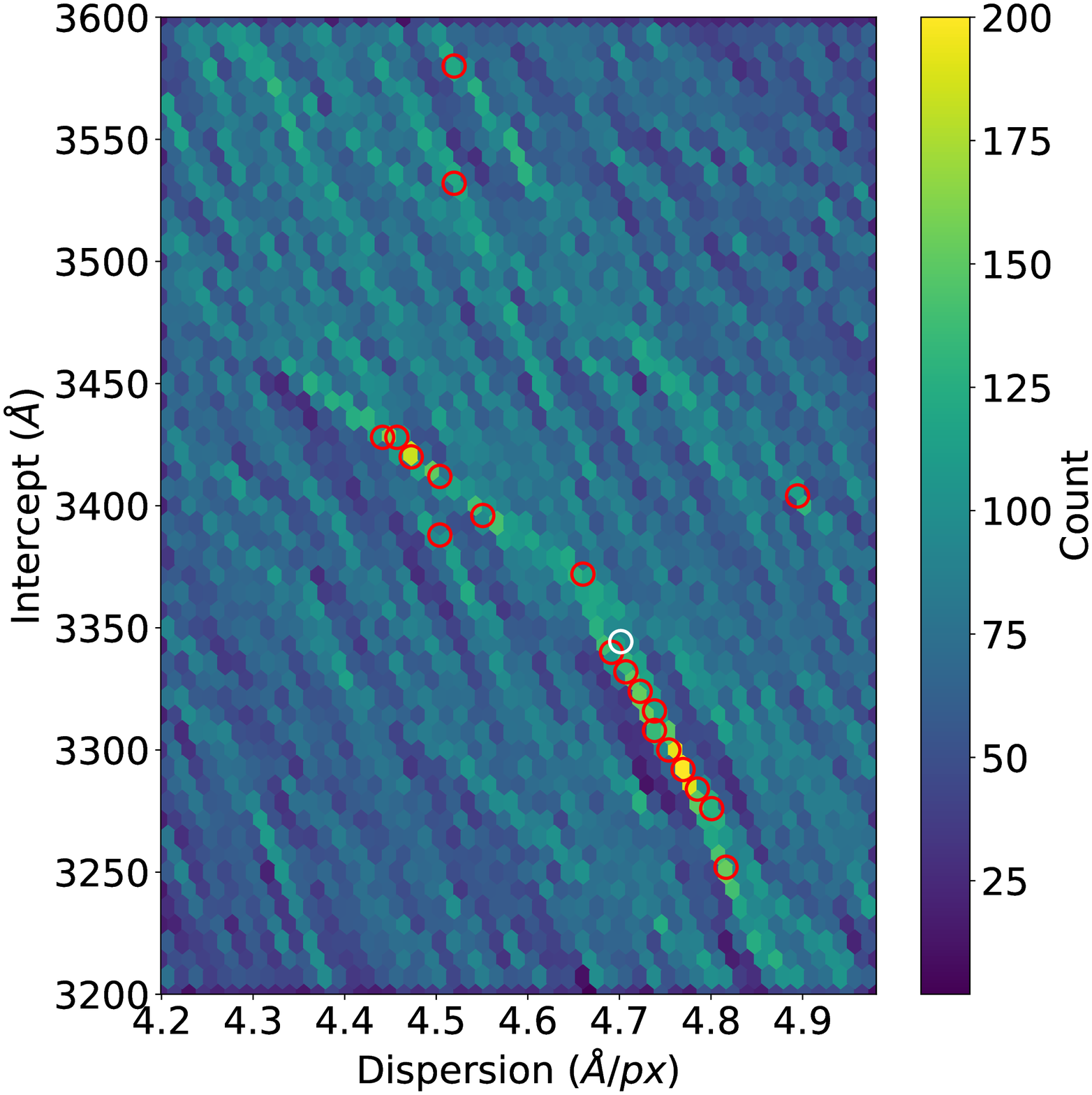}
    \label{fig:hough}
\end{minipage}

\caption{Determining probable correspondences between peaks and emission lines. \textbf{Left:} Cartesian product of all detected peaks in an input spectrum (Xe) and corresponding NIST atlas lines. True correspondences are shown as red stars. The dashed lines mark the top 20 candidate line fits from the Hough transform (right); note that the overdensity follows the true correspondences and is not straight. \textbf{Right:} Hough accumulator matrix. Red circles highlight the top 20 peaks in the histogram. The white circle marks the linear fit to the true correspondences. }
\label{fig:algorithm_plots}
\end{figure}

\section{RASCAL: Method}

Our goal is to find the true line in $A$ for each peak in $P$. Checking all possible sets of pairs of $A$ and $P$ is computationally infeasible, so~\citeauthor{Song:18}'s approach is to search for \textit{plausible} sets of correspondences which are constrained by prior knowledge about the system. Specifically they find solutions which agree with linear approximations to the system ($x_\lambda = Dx + c$). We likely know which lamp was used, and we also have priors on the minimum wavelength ($c$) and dispersion ($D$) of the system. Let $A'$ be a filtered atlas which only contains lines within a user-specified range of interest. We allow a default tolerance of $\pm 200$\AA~to this value. We constrain $D$ based on the number of pixels in the spectrum and the wavelength range.

 Initially, all possible pairs of peaks and emission lines are enumerated (i.e. the Cartesian product $A' \times P$.) An example plot is shown in Figure~\ref{fig:algorithm_plots}. The Hough transform~\citep{hough1962method} is used to search for linear correspondences among these enumerated pairs. The result of this is a histogram of possible lines in $(D, c)$ space, illustrated in Figure~\ref{fig:algorithm_plots} (right). Peaks in the Hough accumulator map (red circles) to lines which pass through or near lots of peak-line correspondences; we call this a candidate set.

The original algorithm suggests fitting models to each candidate set separately and then choosing the best. In our experience, this fails when there is a large amount of curvature in the model function. Instead, we consider the top $N$ candidate sets simultaneously (we set $N = 20$ by default). For each peak, we choose the most common best-fit atlas line from the top candidate sets. This acts somewhat like a piece-wise linear fit and allows us to extract most of the correct matches from both the red and blue regions of the spectrum, shown in~\ref{fig:algorithm_plots} (left). RANdom SAmple Consensus~\citep[RANSAC][]{Fischler:1981:RSC:358669.358692} is used to robustly fit a 4th or 5th order polynomial model to the candidate correspondences. This model is used to return atlas correspondences for each peak, which can be passed to a more sophisticated fitting function e.g.an analytical model of the instrument such as \citep{10.1117/1.OE.52.1.013603}.

\subsection{Implementation and Example Results}

We implemented the algorithm in Python and only require a few common dependencies: \texttt{numpy}~\citep{doi:10.1109/MCSE.2011.37}, \texttt{scipy}~\citep{2019arXiv190710121V}, \texttt{matplotlib}~\citep{doi:10.1109/MCSE.2007.55} and \texttt{astropy} (for unit support)~\citep{astropy:2013, astropy:2018}. The runtime of the algorithm is fast - less than 10 seconds on a Core i5 laptop. We have not performed any serious optimisation for speed. Our code and example \texttt{jupyter} notebooks are available on Github~\footnote{\url{https://github.com/jveitchmichaelis/rascal}.}.

\begin{figure}
    \centering
    \includegraphics[width=0.9\linewidth]{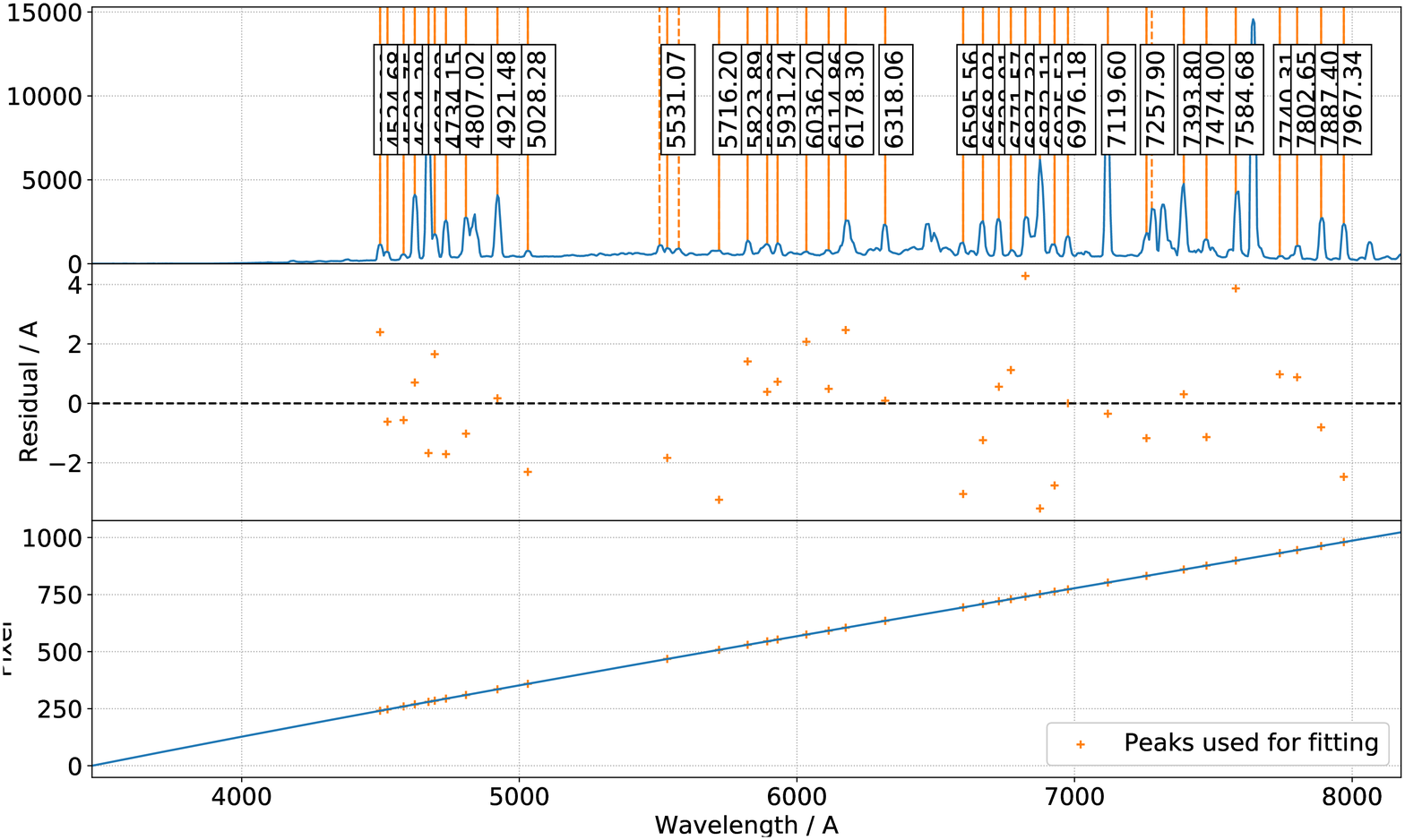}
    \caption{Example calibration using a Xenon arc lamp spectrum from the SPRAT spectrograph on the Liverpool Telescope.}
    \label{fig:results}
\end{figure}

An example of the algorithm applied to a Xenon arc lamp spectrum taken with the SPRAT spectrograph~\citep{2014SPIE.9147E..8HP} on the Liverpoool Telescope~\citep{2004SPIE.5489..679S} is shown in Figure~\ref{fig:results}. Scipy was used for initial peak finding and a wavelength range of 3400-8100 \AA was used for atlas filtering.

\section{Summary}

RASCAL is currently being integrated into the ASPIRED pipeline and we are currently testing the algorithm on spectra from other instruments. We are also performing a more rigorous analysis on suitable default settings that will be applicable to more instruments, and assessing how robust RASCAL is to e.g. spurious peaks/atlases.

\bibliography{P10-37}


\end{document}